# Towards electrically driven organic semiconductor laser with field-effective transistor structure

Thangavel Kanagasekaran[1]*, Hidekazu Shimotani[2]*, Keiichiro Kasai[2], Shun Onuki[2], Rahul D. Kavthe[3], Ryotaro Kumashiro[1], Nobuya Hiroshiba[4], Tienan Jin[5], Naoki Asao[3] and Katsumi Tanigaki[1,2]*

**Affiliations:**

[1]WPI-AIMR, Tohoku University, 2-1-1 Katahira, Aoba, Sendai 980-8577, Japan.

[2]Department of Physics, Graduate School of Science, Tohoku University, 6-3 Aramaki, Aza-Aoba, Aoba-ku, Sendai 980-8578, Japan.

[3]Department of Chemistry and Materials, Faculty of Textile Science and Technology, Shinshu University, 3-15-1 Tokida, Ueda 386-8567, Japan.

[4]Department of Physics, IMRAM Polymer Hybrid Materials Research Center, Waseda University, Japan.

[5]Department of Chemistry, Graduate School of Science, Tohoku University, 6-3 Aramaki, Aza-Aoba, Aoba-ku, Sendai 980-8578, Japan.

*Correspondence to: kanagasekaran@gmail.com, shimotani@tohoku.ac.jp, tanigaki@m.tohoku.ac.jp.

**Abstract:** Laser is one of the most important discoveries in the 20[th] century, and inorganic semiconductor lasers (ISCL) are most frequently used in many applications nowadays. Organic semiconductor lasers (OSCL) have many attractive features when they compared to ISCLs, such as flexibility, human friendliness, feasible and inexpensive production process, light weight, and multicolor emission. However, electrically driven OSCLs (el-OSCL) have not yet been realized, although they are possible in an optically driven mode. Here, we report that an el-OSCL can be realized in field-effect transistor (FET) structure. The FET el-OSCL with distributed feedback (DFB) construction is made using a 5,5"-bis(biphenyl-4-yl)-2,2':5',2"-terthiophene single crystal as a lasing medium electrostatically laminated on a silicon substrate modified with periodically patterned polystyrene. An emergent sharp-linewidth emission spectrum to the resolution limit of a detector and a non-linear increase in intensity above the threshold current density of ca. 1 kA cm$^{-2}$ was observed, being indicative of lasing. Discussions on the possible realization of lasing in el-OSCLs can be made from the comparison between optical and electrical-driven mode.

Laser stands for light amplification by stimulated emission of radiation and the Nobel Prize was awarded for its discovery in 1964[1]. High-performance blue-color lasers missing for long years in the past were realized in GaN[2]. Nowadays lasers are indispensable devices not only in industry but also in daily life. Compared to inorganic semiconductors (ISCs) such as GaN, organic semiconductors (OSCs) have many fascinating features such as flexibility, a feasible process, low cost-high performance as well as human-friendly texture[3]. In addition, considering that there are a variety of molecules in OSCs, lasers showing multi-color emission can be anticipated ranging from nearly deep-ultraviolet to far-infrared, and therefore realization of OSC lasers (OSCLs) has been an important, challenging, and cutting-edge research theme in optoelectronics. The research reported in 2010 by Hendrik Shön and coworkers[4], showing the realization of an OSCL in field effective transistor (FET) structure by employing tetracene (one of the most conventional OSCs) was welcomed with a surprise and a surge of scientific interest from many researchers, and consequently has inspired many future perspective applications.

However, the report has been recognized afterwards to be one of the serious misconducts in science and technology so far ever made in the world, and the paper was withdrawn with having left many arguments. Nevertheless, many researchers have been continuing to target its realization.

Optically driven solid-state OSCLs (op-OSCLs) have later been exemplified, and laser emission with various colors has been demonstrated to date[5-8]. However, any successful operation of electrically driven OSCLs (el-OSCLs) had still not yet been achieved due to the low carrier mobilities and the low electron-injection efficiency in OSCs.[9-11] The carrier injection exceeding 10 kA cm$^{-2}$, which is presently considered to be required, was indeed far apart from the injected current level to be accessible in 2010. In the approaches towards the realization of el-OSCLs, one can find three intriguing reports[12-14], but any of them are commented not to afford to provide firm experimental evidence for operation in the electrically driven (el-driven) mode[15]. For example, Yokoyama et al. approached to an el-OSCL by targeting on 4,4'-bis((N-carbazole)styryl)biphenyl showing high quantum efficiency of emission[12], but the reported current density of $J = 0.4$ A cm$^{-2}$ is apparently too small to realize el-OSCLs. Bisri et al.[12,16] tried to make an el-OSCL by employing 5,5"-bis(biphenyl-4-yl)-2,2':5',2"-terthiophene single crystal (sc-BP3T) in a bilayer structure with high current injection and a spectrum narrowing was observed, but neither clear threshold current density ($^{el}J_{th}$) nor non-linear increase in emission intensity ($I_{em}$) was observed. The other claim of realization of el-OSCLs can be found for a microcavity resonator with dielectric mirrors[14]. However, this report was neither lasing nor amplified spontaneous emission (ASE), because the $^{el}J_{th}$ of 0.86 A cm$^{-2}$ and the spectrum narrowing as well as the increase in intensity of emission above $^{el}J_{th}$ is too small to be justified from the common understanding of el-OSCLs.

Recently, we reported, as the new concept of electrodes[17], the ambipolar carrier injection exceeding $^{el}J = 25$ kA cm$^{-2}$ in a rubrene single-crystal FET, which importantly exceeds the $^{el}J_{th}$ of 10 kA cm$^{-2}$ considered to be required for lasing in the FET structure of sc-BP3T[18,19].

Accordingly, we decided to try to make the realization of el-OSCLs in this stage. Here, we report intriguing indication of a successful el-OSCL with FET structure by employing distributed feedback (DFB) construction using sc-BP3T as a laser medium, where a very sharp laser emission reaching the resolution limit of a detector evidently emerges with a nonlinear increase in its intensity above the clear $^{el}J_{th}$ at around 1kA cm$^{-2}$.

Laser is an electronic device that can be composed of a laser medium (a light emitting material) and a resonator (a gain medium with an optical feedback system). In the case of OSCLs, two types of feedback systems can generally be considered, one being a Fabry-Pérot resonator (FPR) and the other being a DFB construction. We fabricated an el-OSCL with FET structure using the DFB nano construction as shown in Fig. 1, since DFB can reduce $^{el}J_{th}$. Moreover, the laser emission with extremely narrow spectrum linewidth (FWHM, full width at half maximum) can experimentally be detected, which can easily be differentiated from other emissions to provide the firm evidence of whether laser is realized. We observed both a very sharp laser emission spectrum to the resolution limit of an employed detector and a nonlinear increase in emission intensity with the clear $^{el}J_{th}$ as described in the next paragraph.

Below the $^{el}J_{th}$ of ca. 1 kA cm$^{-2}$, DFB-el-OSCLs of sc-BP3T still showed a broad emission spectrum arising from the vibronic transitions indexed to be 0-1, 0-2, and 0-3 from the exitonic singlet state to the vibronic ground states in addition to the very weak optically forbidden 0-0 emission. This indicates that the DFB construction gives only a partial influence around the bottom nano-structured patterns among the total emission below $^{el}J_{th}$. The emission spectrum was very similar between the el- and the op-driven mode below $^{el}J_{th}$ and $^{op}J_{th}$. When larger current beyond the $^{el}J_{th}$ was injected, a very narrow spectrum reaching the resolution limit of a 2 nm-detector to be markedly differentiated from the broad emission immediately emerged with a sharp nonlinear increase in its intensity as shown in Fig. 2. A similar change was observed at $^{op}J_{th}$ = 2 µJ cm$^{-2}$ in the DFB-op-OSCL. In the DFB-op-OSCL, the intensity of the emergent narrow spectrum became greatly enhanced when light fluence increased further up to 39 µJ cm$^{-2}$ far above the $^{op}J_{th}$, and then correspondingly the broad emission spectrum relatively became quite small as shown in the inset of Fig. 2 (upper left panel). The FWHM of the emergent narrow spectrum was 2 nm to the resolution limit of a detector with a resolution of 2 nm both in the op- and the el-driven mode. We measured the FWHM of DFB-op-OSCL using a higher resolution detector of 0.2 nm and the FWHM was measured to be 0.22 nm as shown in the inset of Fig. 2 (upper right panel). In the case of DFB-el-OSCL, however, the current injection was not able to be further increased beyond 1.8 kA cm$^{-2}$ most probably due to the strains given when BP3T single crystals were laminated on the nano patterned PS with DFB structure. These results indicate that lasing is realized in both the DFB-el-OSCLs and DFB-op-OSCLs.

The important discussion to be made is whether the real lasing takes place by stimulated amplification of light via a repeated optical feedback process in our el-OSCLs. Lasing is a special physical phenomenon, which can be realized when light emits with sufficiently high intensity, accumulated, and resonantly intensified in a multiple coherent process by stimulated emission in a gain medium of light between the two reflectors so called as a resonator. In order to confirm the lasing by differentiating it from other similar phenomena, the following experimental evidence is generally required. (1) A clear $J_{th}$ to be most commonly viewed in the logarithmic plot of $I_{em}$ vs. $J$, (2) significant spectral narrowing above $J_{th}$, hopefully approaching to the diffraction limit within the resolution of a detector and (3) evidence of the optical feedback in a resonator.

One of the most confusing events with lasing is ASE, where spontaneously emitted photons are coherently amplified in a single pass inside a gain medium, and similar properties to those of lasing can frequently be acquired in ASE. The intrinsic dissimilarities between lasing and ASE are whether light amplification is made through a single pass (ASE) or a multiple pass (laser) via a feedback process inside a lasing medium, the latter of which can be proven under the existence of a resonator with two opposite reflection mirrors (FPR) or DFB via the Bragg reflections in the periodic nano structure. Consequently, laser or ASE can be justified by the evidence of the repeated multiple feedback process in a resonator. The firm evidence in the case of DFB can be given by the spectrum dip around at $\lambda=2n_{eff}\Lambda$ ($\lambda$: wavelength of the emission, $n_{eff}$: the effective refractive index, $\Lambda$: the period of DFB grating) due to the Bragg reflections prior to entering the lasing mode.

Looking back our experimental results on DFB-OSCL in the el- and the op-driven mode, a spectrum dip was clearly viewed around at $\lambda = 615$ nm, which is expected given the conditions of $n_{eff} = 2.2$ and $\Lambda = 140$ nm. Furthermore, a greatly sharp emission notably to be differentiated from the normal one immediately emergees above the $^{el}J_{th=}$ 1 kA cm$^{-2}$ in the el-mode and the $^{op}J_{th}=$ 2 µJ cm$^{-2}$ in the op-mode together with significant nonlinear increase in its integrated intensity, respectively. The spectrum narrowing comes to 2 nm in the resolution limit of our detector for both op- and el-driven mode. In the case of DFB-op-OSCL, the accurate FWHM can further be confirmed to be at least 0.22 nm by using a higher resolution detector of 0.2 nm resolution. These are the important first experimental evidence to confirm that we have come to the realization of laser in DFB-el-OSCLs with FET structure. It is noted, however, that a similar high intensity lasing to that of DFB-op-OSCL was not fulfilled in DFB-el-OSCL, since the current density cannot be raised sufficiently above the $^{el}J_{th}$ due to the inferior mobility of BP3T laminated on the substrate surface with the DFB nano structure of PS as we described earlier.

As described earlier in the case of DFB, it is shown that lasing occurs in both el- and op-driven modes. We made similar experiments on FPR-el-OSCLs and also observed both a nonlinear increase in $I_{em}$ and its spectral narrowing above the $^{el}J_{th}$ of 10 kA cm$^{-2}$, which is about one-order larger than 1 kA cm$^{-2}$ observed in the case of DFB-el-OSCL. The situation is displayed together with a change of the emission spectrum in Fig. 3. Intriguingly, by changing the film thickness of BP3T single crystals, the ratio of the 0-1 emission relatively increases compared to the 0-2 emission as the film thickness (*d*) decreases. This is because the confinement of electromagnetic waves is dependent on the film thickness. Consequetly, selective light emission of either the 0-1 or the 0-2 can be manupilated by changing *d*. In FPR-el-OSCL, however, the increase in $I_{em}$ stopped at an intermediate amplification process with a factor of 3 to 5 times for the 0-1 emission and 10 times for the 0-2 emission. Being associated with the cease of amplification at the intermediate stage, spectral narrowing also stopped at around 7 to 8 nm and did not come to the resolution limit of a detector with 2 nm. It is noted that this situation of FPR-el-OSCL is a little bit different from what was observed for DFB-el-OSCL. In the FPR-el-OSCL, we were able to increase the input current density sufficiently above the $^{el}J_{th}$ but the lasing stopped in the intermediate situation, whereas in the DFB-el-OSCL we were not able to provide sufficient $^{el}J$ above the $^{el}J_{th}$ althtough the lasing continues to proceed towards the same ideal lasing realized in DFB-op-OSCL. Arguments of whether lasing occurs or not in the case of FPR-el-OSCL will be made later in the discussion section.

**Discussion**

In order to discuss the lasing realized in the DFB-el-OSCLs, we first compare the exciton densities ($N(ex)$) populated between the el- and the op-driven modes.

In the op-driven mode, we pumped the laser medium using an N$_2$ pulse laser (the wavelength $\lambda = 337.1$ nm, pulse width $w = 3.7$ ns). The exciton density $^{op}N(ex)$ generated in the op-driven mode can be estimated by $^{op}N(ex) = {}^{op}J_{th}\lambda/hc$, where *h* and *c* are the Planck constant and the velocity of light. We took two corrections into consideration for having the realistic estimates; one being the decay of emission (the singlet fluorescence lifetime ($\tau$) is reported to be 1.68 ns for BP3T[20]) to be made using the equation of $^{op}N(ex)(\tau/w)[1 - \exp(-w/\tau)]$. The other one is the absorption of incident light through three-dimensional sc-BP3T using absorption coefficient $\alpha = 1.016 \text{ μm}^{-1}$ by employing the equation of $^{op}N(\text{ex})[(1 - \exp(-\alpha d))/d]$, where $d = 270$ nm. Consequently, $^{op}J_{th}$ in the op-driven mode gives $^{op}N_{th}(\text{ex}) = {}^{op}J_{th}(\lambda/hc)(\tau/w)[1 - \exp(-w/\tau)] [(1 - \exp(-\alpha d))/d]$. Applying the equation, $^{op}N_{eh}(\text{ex})$ can be estimated to be $2.6 \times 10^{17}$ for DFB-op-OSCL ($^{op}J_{th} = 2$ μJ cm$^{-2}$).

In the el-driven mode, the populated singlet exciton density at the threshold current density can be estimated by $0.25\tau^{el}J_{th}/(eW_{RZ})$. Here $e$ is the elemental charge, $W_{RZ}$ is the recombination zone width to create excitons (generally ca. 1 - 3μm) and the factor of 0.25 is the ratio of singlet excitons formed by the recombination of electrons and holes. From the above relationship, $^{el}J_{th} = 1.0$ kA cm$^{-2}$ for DFB-el-OSCL gives singlet exciton density $^{el}N_{th}(\text{ex})$ to be $2.6 - 7.2 \times 10^{16}$ cm$^{-3}$. Although these evaluated values of exciton density may sensitively depend on the employed physical parameters, the values will not be far apart from those of the real situations. From these quantitative discussions, it seems that $^{el}N_{th}(\text{ex})$ required for lasing in the el-driven mode at the threshold is a little bit larger than $^{op}N_{th}(\text{ex})$ in the op-driven mode.

The situations described here are displayed as the comparison between DFB and FPR in Fig.4(A). This would be very reasonable for considering the quenching of excitons, such as electric field given by both electrodes, charge of carriers, and singlet-triplet annihilations in the case of el-driven OSCLs.

In order to figure out which is realized, laser or ASE, in FPR-OSCLs, we made simple but important experiments on BP3T by comparing between a good single crystal with two opposite mirror facets (sc-BP3T) and the same single crytal with one facet mirror plane mechanically damaged (dm-BP3T) as shown in Fig.4(B) in the op-driven mode. Since we could observe longitudinal modes depending on the cavity length ($L$) as the evidence of laser in the case of op-driven mode, non linear increase in $I_{em}$ above $^{op}J_{th}$ is evidently caused by a multiple feedback amplification process of lasing for sc-BP3T with two opposite mirror facets.

On the other hand, only one-way stimulated amplification of ASE is possible in dm-BP3T because one mirror facet was mechanically damaged and no multiple feedback is possible. Although a nonlinear increase was observed at a similar $^{op}J_{th}$ within the errors of our present experiments, the gradient of the $I_{em}$-$^{op}J$ plot for dm-BP3T above the $^{op}J_{th}$ was significantly lower than that of sc-BP3T. We also plotted the $I_{em}$-$^{el}J$ correlation of our FPR-el-OSCL in Fig.4(B) by adjusting the $^{el}J_{th}$ to the $^{op}J_{th}$, and found that the slope of $I_{em}$-$^{el}J$ for FPR-el-OSCL is alomost the same as or a little bit grearter than that of sc-BP3T in the op-driven mode. These experimental data strongly suggest that the emission observed in our FPR-el-OSCL could be lasing but not ASE.

In FPR, various emissions in wavelength are permitted under lasing corresponding to the standing waves possibly existing in a resonator, and the laser emission approaches to the higher quality laser emission as well as the narrower FWHM with better time and space coherency as the emission intensity increases away from $J_{th}$. According to the relationship of $\Delta\lambda = \lambda^2/2n_gL$, where $\Delta\lambda$ is the separation of longitudinal modes, $n_g$ the group refractive index of the lasing medium, and $L$ is the resonator cavity length. For the 0-1 emission FPR-op-OSCL of $\lambda = 575$ nm, $n_g = 4.6^{21}$, and $L = 35$ μm and the 0-2 emission FPR-op-OSCL of $\lambda = 616$ nm, $n_g = 3.7^{21}$, and $L = 17$ μm, of the estimated $\Delta\lambda$ of 1.0 and 3.0 nm, respectively, are comparable to the observed value of 0.6 and 2.6 nm, respectively.. However, the present our experiments show that lasing stopped under intermediate lasing by some reasons in the case of FPR-el-OSCLs as described earlier.

For reaching a similar good lasing condition in FPR-el-OSCLs to that in FPR-op-OSCLs, the following factors should deeply be taken into account. First, the temperature increase during the lasing process in the el-driven mode may be important. Because a large current density of ca. 10 kA cm$^{-2}$ is injected into FPR-el-OSCLs with high electrical resistivity, innegligible Joule heating is generated and the lasing process may become unstable. We evaluated using a simulation based on a thermal diffusion process that temperature increase would be smaller than ca. 11 K in the room temperature operation, and such temperature increase may give less stability under the lasing in the vicinity of $^{el}J_{th}$ because of the temperature dependent loss in a resonator. Another factor to be considered would be the narrow recombination zone in el-OSCLs with FET structure, which corresponds to the stripe width in the case of gain-guiding multimode ISCLs[22]. In our el-OSCLs, the recombination zone is as narrow as at most several μm, which is greatly smaller than optically excited zone of 149 μm in op-OSCL. The confinement of exciton and photons may also be important in FRP-el-OSCLs, although this is of less significance in the case of DFB-el-OSCL with nano structure feedback system. The other important consideration should be made for the quenching of singlet excitons by triplet excitons (singlet-triplet (ST) annihilation), scattering by charge carriers, and the influence of the metal electrodes. Among these, the most apparent influence on the singlet exciton quenching may be the ST annihilations, which have experimentally been reported and discussed. A comment was reported based on pthototransient spectroscopy that a pulse excitation shorter than a few handred nano second is necessary considering the ST annihilation for preventing lasing and the el-driven laser cannot be realized in the cw mode in OSCLs composed of aluminum-quinoline type compounds[24].

Importantly, when we see the experimental data carefully, the slope of $I_{em}$ as a function of $J$ in FPR-el-OSCL is lower than that of FPR-op-OSCL below $J_{th}$ as seen in Fig.3 and Fig. 4(B). Because a larger amount of triplet excitons are directly generated in the el-driven mode than that in the op-driven mode, the ST annihilations should seriously be taken into consideration in the el-driven mode. The $I_{em}$ of the normal emission below $^{el}J_{th}$ of 10 kA cm$^{-2}$ is not ideally proportional to the $^{el}J$ and its slope becomes rather low showing lower efficiency when $^{el}J$ is large. The quenching of singlet excitons actually occurs, but we showed that the lasing still occurs in DBF-el-OSCL and FPR-el-OSCL with FET structure of BP3T at around 1 kA cm$^{-2}$ and at 10 kA cm$^{-2}$, respectively, in the cw mode. Although improved experiments are still necessary especially in the case of FPR-el-OSCL, more sophisticated discussions would be necessary in the future.

**Conclusion**

Through continuous efforts over many years after the scientific misconducts reported in 2000, we showed the important experimental indications that lasing of OSCLs with FET structure in the el-driven mode can become possible and an OSCL can reach the intrinsic lasing under the el-driven mode for DFB/FPR-el-OSCLs in the cw mode. This is the far-beyond intriguing advancement in the research of OSCLs during the past two decades. We made various important discussions on how lasing can be realized in OSCLs in the el-driven mode. Both intrinsic nonlinear amplification of the emission intensity and the clear threshold current density, in addition to the emergence of a very sharp emission peak given by the optical feedback of the Bragg reflection to be evidenced by the spectrum gap appearing prior to lasing, give the strong indication that we realized the lasing in DFB-el-OSCLs with FET structure. In the FPR-el-OSCLs, lasing starts but stoppes under intermediate condition before reaching the same ideal lasing as that in FPR-op-OSCLs.

The definition of lasing in the text book is the light emission coherently amplified by the stimulated emission of radiation process with repeating optical feedback in a resonator. When optical feedback in a resonator is missing, one-way stimulated emission, ASE, is observed if the gain overcomes the loss in an optical medium. However, different criteria of lasing are frequently given and they depend on the scientists. Even though the lasing defined in the textbook is realized, some of scientists may not say that lasing is realized. For such scientists, the emission with high quality of time and space coherency is required for calling the emission as laser. The definition of lasing is not unique in this sence. Although we claim in this manuscript that lasing was evidenced in DFB-el-OSCLs according to the definition of a text book, this may not be permitted by some other scientists.

Further scientific discussions on the lasing mechanism and technological efforts for improving the lasing characteristics in OSCLs in the el-driven mode will be required for pushing them up to the laser operation comparable with that in the op-driven mode in the future. Since markedly different advantages from those of silicon and III-V compound semiconductors can be considered for el-OSCLs and they provide a simple, cheap, wavelength-tunable, and flexible laser-light source, a new chapter for optoelectronics may open in the future.

## Acknowledgments

The authors thank Teruya Ishihara, Hiroyuki Yokoyama, and Masayuki Yoshizawa for their fruitful discussions. This work was supported by JSPS KAKENHI Grant Number JP17H05326 and 18H03883. K.T. acknowledges JSPS KAKENHI Grant Number 17H05326 and 18H04304. H.S. acknowledges JSPS KAKENHI Grant Number JP24684023, JP25610084, and JP16K13826, JGC-S Scholarship Foundation, and CASIO Science Promotion Foundation. The work was also supported by the World Premier International Research Center Initiative (WPI) from MEXT of Japan and International collaborative bilateral-country program by JSPS.

## Author contributions

The research idea and design of experiments for its confirmation were made by T. K., H. S., and K. T. BP3T was synthesized by R. D. K., T. J., and N. A. FET experiments were performed by T. K. and optically driven mode experiments were made by S. O. and K. K. The experimental data were analyzed, and the manuscript was prepared by T. K., H. S., and K. T..

## Competing interests

Authors declare no competing interests.

## Methods

We compared performances of various OSCs as a laser medium for realization of el-OSCLs in a FET structure[17, 25-28], and selected sc-BP3T as the best candidate for realizing an el-OSCL, because it has high mobility of both holes and electrons as well as high internal quantum efficiency of light emission[16].

**Single-Crystal Growth.** BP3T single crystals were grown by a physical vapor transport method under Ar gas flow from pristine BP3T powder in a tube furnace with temperature gradient. The temperature gradient of the furnace was from 265 to 385 °C, and the Ar gas flow rate was maintained at 42 sccm during the crystal growth process. High quality single crystals were obtained with different shapes depending on the position of the crystal growth zone. We selected the high-quality crystals among them and carried out both electrically- and optically-driven experiments.

**Optical measurements.** Photo-pumped measurements were performed by irradiating a laser light using a nitrogen nanosecond-pulsed laser at 337.1 nm (Stanford Research Systems, Inc., NL100). The laser spot was an ellipse with a long axis of 746 μm and a short one of 149 μm at the $1/e^2$ intensity. The incident angle to the crystals was 30°. The pulse fluence was controlled by neutral density filters. Light emissions were detected in the parallel direction to the ab-plane of crystals using a high-sensitivity optical fiber spectrometer with wavelength resolution < 2 nm (Hamamatsu Photonics K.K., PMA-12 Photonic Multichannel Analyzer C10027-01) or a high-resolution fiber spectrometer with wavelength resolution of 0.2 nm (Ocean optics, Inc., HR2000+).

**Device Fabrication.**

FPR-el-OSCLs were also fabricated in the bottom-gate and top-contact FET configuration. A $p^{++}$-Si substrate with a $SiO_2$ layer of 300 nm in thickness was spin-coated with a polystyrene thin film from 1wt% solution in toluene. The thickness of the polystyrene film was 25 nm. The substrate was dried in Ar atmosphere at 75 °C for more than one night (~15 hours), followed by lamination of a BP3T single crystal on it. Single crystals with parallel edges were selected in order to prepare devices. For electrodes, first we deposited CsF (1.0 nm) to enhance an electron injection efficiency followed by deposition of Ca and then Au was deposited by employing a stencil mask with a vacuum thermal evaporation method.

**Device characterization.** All electrical and optical measurements were carried out with a B1500A Semiconductor Device Parameter Analyzer (Keysight Technologies Inc.) and an optical microscope combined with a CCD camera and a spectrometer in a glove box under Ar gas atmosphere at room temperature. The microscope was tilted by 0° or 80° to the c-axis to observe the light emission from the top or the edge of a single crystal.

**OSCLs of BP3T with DFB construction**

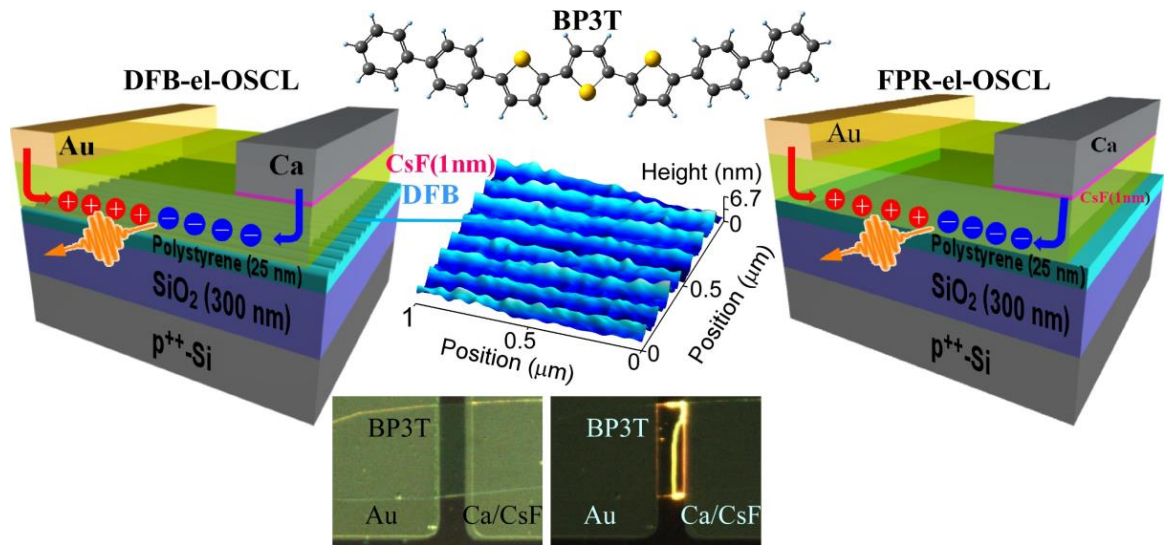

**Fig. 1| DFB- and FRP-el-OSCL devices in electrically driven mode operation and laser emission.** Structure of electrically driven organic semiconductor laser of BP3T (laser medium) with DFB construction of nano-textured polystyrene (DFB-el-OSCL) and with naturally-fabricated Fabry-Pérot resonator of BP3T single crystal (FPR-el-OSCL) in FET structure with heteroelectrodes of Ca/CsF (for electron injection) and Au (for hole injection). An atomic-force micrograph of the surface structure of a DFB polystyrene layer is displayed. A bird's-eye viewed optical micrograph of DFB-el-OSCL is shown at the bottom and light emission is exibitted.

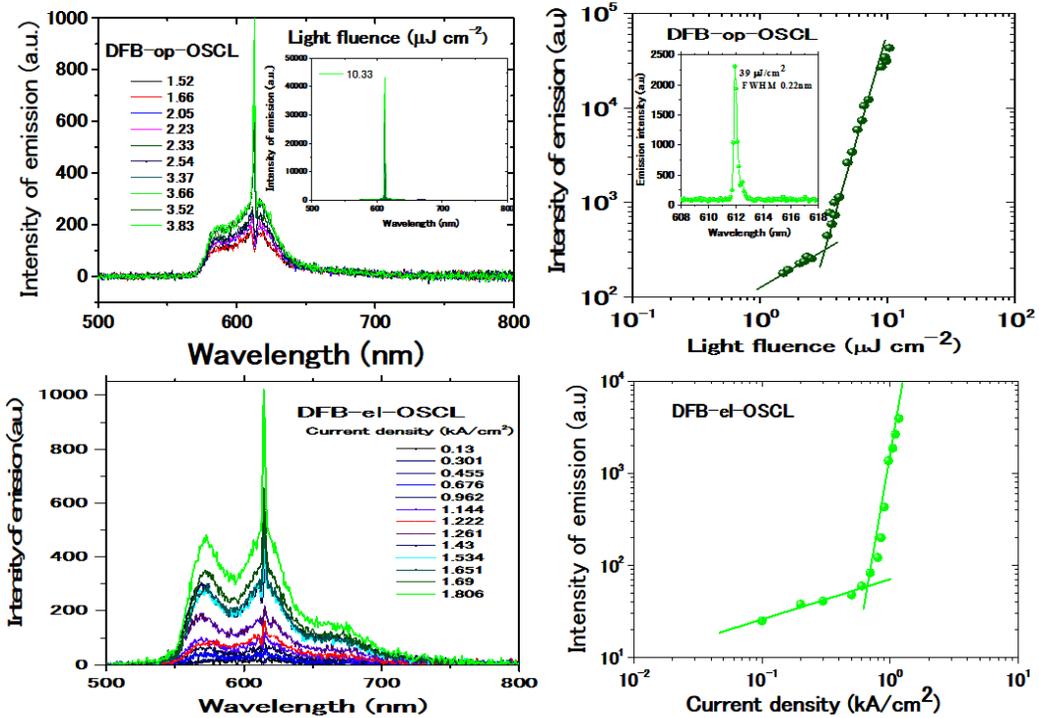

**Fig. 2| Laser emission spectra observed in DFB-OSCL and relationship between emission intensity v.s. current density /light fluence in both el- and op-driven mode.** A very sharp emission with nonlinear increase above the threshold is clearly observed for both el- and op-driven mode. The evidence of existing optical feedback via Bragg reflections from the DFB structrure is observed as the spectrum dip (Bragg dip) prior to lasing as described in the text. FWHM reaches the resolution limit of a detector of 2 nm. Explanded view of emission spectrum in DFB-op-OSCL to be measured by a high resolution detector of 0.2 nm resolution is shown in the inset (upper light panel), where the background normal emission is relatively suppressed due to the storng laser emission intensity and only a very sharp emission with a linewidth of 0.22 nm is evident.

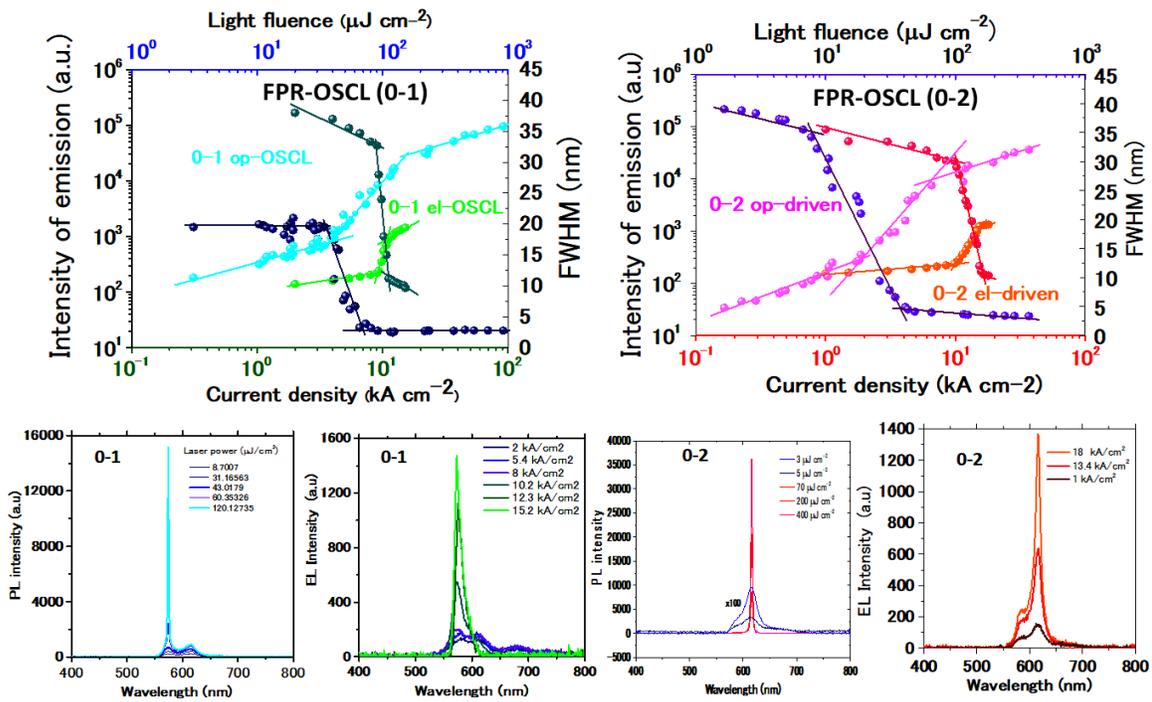

**Fig. 3| Relationship of emission intensity ($I_{em}$) and spectrum width (FWHM) v.s. current density ($^{el}J$)/light fluence ($^{op}J$) and spectrum shape in FPR-OSCLs.** Either 0-1 or 0-2 emission can be selectively lased by changing the film thickness (d) of sc-BP3T. Detailed explanations and discussions are given in the text.

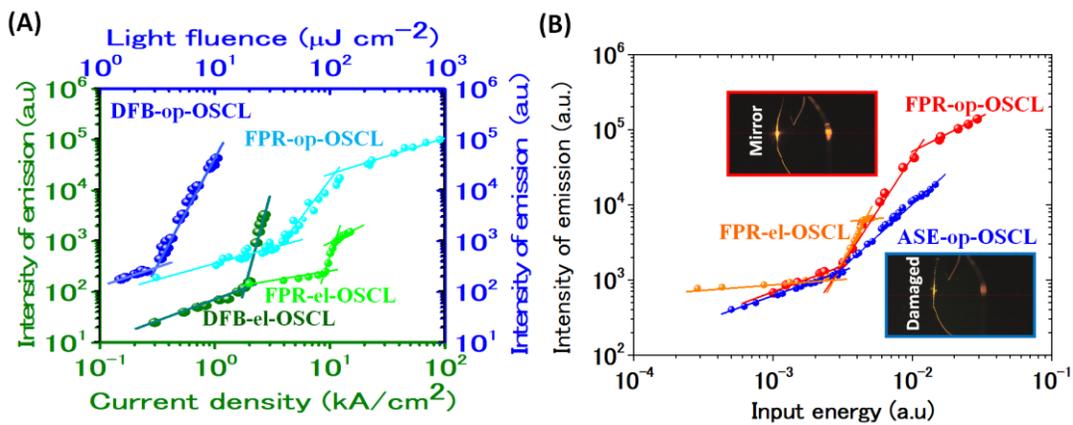

**Fig. 4| Comparsion of emission intensity v.s. input energy betwee DFB- and FPR-OSCL both in el- and in op-driven mode, and comparsion between lasing and ASE in FPR-op-OSCLs.** (A) Intensity of emission as a function of current density or light fluence. (B) Comparsion of relationship of intensity of emission v.s. light fluence between ASE and laser in the op-drivn mode. Nonlinear amplification is markedly differentiated beteen lasing (sc-BP3T with two good mirror fascets) and ASE (BP3T with one mirror fascet mechanically broken). Note that the dependence of intensity of emission as a function of current density observed for FPR-el-OSCLs is almost identical to that of FPR-op-OSCL in the lasing mode as described in the text.